Encoding spatiotemporal asymmetry in artificial cilia with a ctenophore-inspired soft-robotic platform.

David J. Peterman and Margaret L. Byron

Department of Mechanical Engineering, Penn State University, State College, PA 16801, USA

Corresponding author email: djp6286@psu.edu

Key words: artificial cilia, intermediate Reynolds numbers, fluid mechanics, biomechanics, locomotion, zooplankton, magnetic elastomers, soft robotics, ctenophore, spatiotemporal asymmetry

**Abstract:**

A remarkable variety of organisms use metachronal coordination (i.e., numerous neighboring appendages beating sequentially with a fixed phase lag) to swim or pump fluid. This coordination strategy is used by microorganisms to break symmetry at small scales where viscous effects dominate and flow is time-reversible. Some larger organisms use this swimming strategy at intermediate scales, where viscosity and inertia both play important roles. However, the role of individual propulsor kinematics—especially across hydrodynamic scales—is not well-understood, though the details of propulsor motion can be crucial for the efficient generation of flow. To investigate this behavior, we developed a new soft robotic platform using magnetoactive silicone elastomers to mimic the metachronally coordinated propulsors found in swimming organisms. Furthermore, we present a method to passively encode spatially asymmetric beating patterns in our artificial propulsors. We investigated the kinematics and hydrodynamics of three propulsor types, with varying degrees of asymmetry, using Particle Image Velocimetry and high-speed videography. We find that asymmetric beating patterns can move considerably more fluid relative to symmetric beating at the same frequency and phase lag, and that asymmetry can be passively encoded into propulsors via the interplay between elastic and magnetic torques. Our results demonstrate that nuanced differences in propulsor kinematics can substantially impact fluid pumping performance. Our soft robotic platform also provides an avenue to explore metachronal coordination at the meso-scale, which in turn can inform the design of future bioinspired pumping devices and swimming robots.

# 1. Introduction

Motile cilia are ubiquitous flexible structures that play important roles in a variety of functional processes. These hair-like organelles interact with fluids across a range of scales (from microns to millimeters) and are often involved in fluid pumping (Sleigh 1989; Alberts et al. 2002; Sensenig et al. 2009; Chateau et al. 2017; Byron et al. 2021), particle transport (Sleigh 1989; Colin et al. 2010; Ding and Kanso 2015; Gilpin et al. 2017), and locomotion (Brennen and Winet 1977; Craig and Okubo 1990; Tamm 2014; Goldstein 2015). Ctenophores (comb jellies; Fig. 1a) operate at the larger end of the scale, bearing cilia up to a few millimeters in length (Afzelius 1961; Tamm 2014; Heimbichner Goebel et al. 2020). These cilia are bundled into paddle-shaped propulsors (ctenes), which are organized into eight meridional rows surrounding the body. Ctenophores offer unique views into aquatic locomotion because of the range of scales at which they operate (Barlow et al. 1993), their propulsor flexibility and coordination (Herrera-Amaya et al. 2021), and the near-omnidirectional maneuverability provided by the organization of these appendages (Craig and Okubo 1990; Matsumoto 1991; Barlow et al. 1993; Tamm 2014; Herrera-Amaya and Byron 2023). These properties make ctenophores an interesting target for bioinspired technologies, ranging from fluid-pumping devices to aquatic robots.

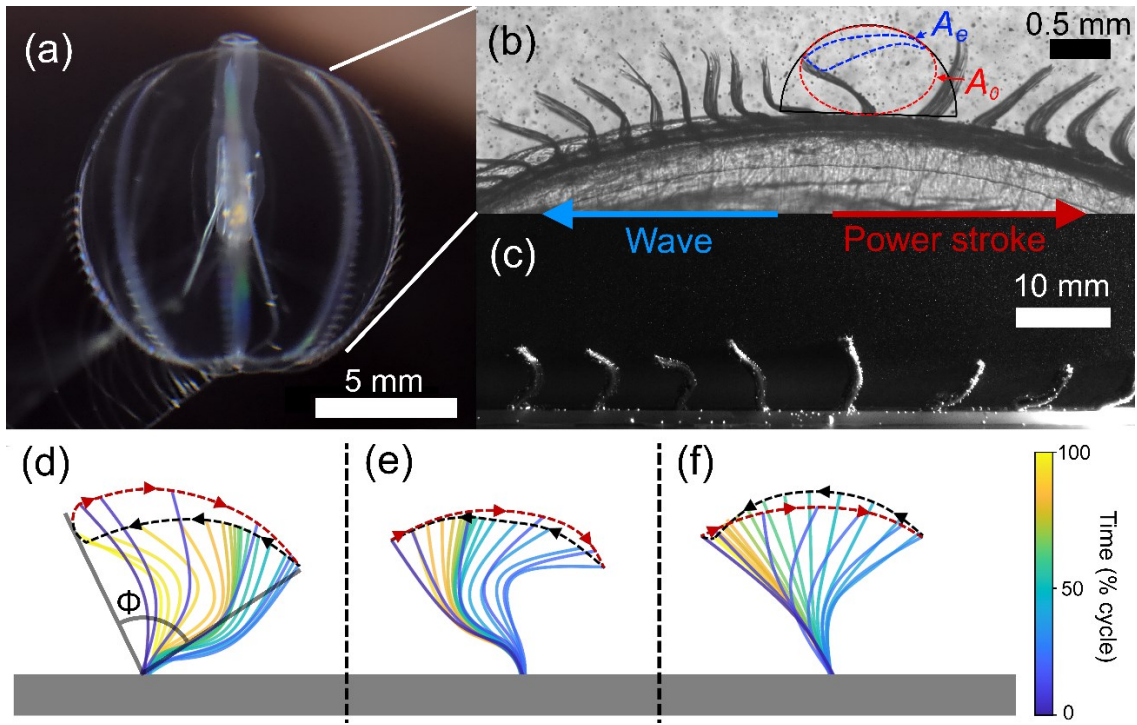

Figure 1: (a) Example of a ctenophore (*Pleurobrachia bachei*) bearing eight rows of propulsors (ctenes). (b) Single row of propulsors (ctene row). The blue dashed line represents the location of the ctene tip throughout its beat cycle, enclosing the area $A_e$. The red dashed line shows the area, $A_0$, of an ellipse inscribed within the maximum possible reach of the ctene (black half circle). These two areas are used to characterize spatial asymmetry as $Sa \equiv A_e/A_0$ (Herrera-Amaya et al. 2021). (c) Soft robotic, ctenophore-inspired platform with plate-like propulsors composed of magnetic silicone elastomers. (d-f) Midline kinematics of the three propulsor morphologies investigated here (convex, concave, and flat, respectively). Arrows denote the motion of the propulsor tips over one beat cycle (red and black portions show the trajectory of the propulsor tips during the power stroke and recovery stroke, respectively). Φ = stroke amplitude.

      Organisms swimming in a viscous-dominated regime must coordinate their propulsors in a way that breaks symmetry (Purcell 1977; Michelin and Lauga 2010; Takagi 2015). One way to incorporate asymmetry is sequentially beating propulsors with a phase lag between neighbors –

i.e., metachronal coordination (Byron et al. 2021). For swimming and pumping, these collective beating patterns are often coordinated as a wave that travels opposite to the power stroke direction (i.e., antiplectic wave; Fig. 1b). Antiplectic metachronal waves generate fundamentally different fluid flow compared to synchronous beating (Khaderi et al. 2011; Takagi 2015) and improve the efficiency and steadiness of flow (Craig and Okubo 1990; Alben et al. 2010; Larson et al. 2014; Chateau et al. 2017; Ford and Santhanakrishnan 2021).

The kinematics of individual propulsors can provide further asymmetry via non-reciprocal motion. Generally, the propulsor extends during the power stroke (increasing thrust) and collapses during the recovery stroke (reducing drag), thereby increasing spatial asymmetry (Khaderi et al. 2010; Herrera-Amaya et al. 2021; Peerlinck et al. 2023). Additionally, increasing the speed of the power stroke relative to the recovery stroke increases temporal asymmetry, which increases net flow at scales where inertial effects are important (Gauger et al. 2009; Khaderi et al. 2012; Semati et al. 2020; Herrera-Amaya et al. 2021). Both metachrony and non-reciprocal motion (spatial asymmetry) are used at lower Reynolds numbers (Re<<1) by ciliated microorganisms (Lauga and Powers 2009; Michelin and Lauga 2010). At low Re, flow is time reversible and spatially symmetric motion of a single propulsor does not generate net fluid displacement, even if temporal asymmetry is present (Purcell 1977). However, metachronal coordination and spatially asymmetric beating kinematics can still provide benefits at larger scales (1<Re<10000), from increased thrust production to synergistic fluid interactions generated by neighboring propulsors (Lim and DeMont 2009; Murphy et al. 2011; Ford et al. 2019; Garayev and Murphy 2021; Santos et al. 2023). Ctenophores (along with many crustaceans, polychaetes, and other animals) operate within the intermediate Re range (Barlow et al. 1993; Vogel 2008; Murphy et al. 2011; Byron et al. 2021; Daniels et al. 2021; Herrera-Amaya et al. 2021; Lionetti et al. 2023), presenting an opportunity to investigate the efficacy of metachronal coordination when neither viscous nor inertial effects can be neglected (Vogel 2008; Klotsa 2019; Derr et al. 2022).

Bioinspired robotic models can elucidate the complex relationships between propulsor kinematics, morphology, scale, coordination, and hydrodynamics. Furthermore, soft robotics and smart materials can be used to more closely match the flexibility of biological structures, particularly relative to traditional rigid models. Of interest for the study of metachronal coordination are several recently developed techniques for the fabrication of artificial cilia (Gu et al. 2020; Zhang et al. 2020, 2021; Islam et al. 2022; Peerlinck et al. 2023). These artificial cilia are actuated via pneumatics, electric stimulation, photo-responsive materials, acoustic action, and other methods (for reviews, see (Zhang et al. 2021; Islam et al. 2022; Sahadevan et al. 2022; Peerlinck et al. 2023)). Here, we focus on magnetic actuation, where artificial cilia are composed of magnetoactive materials—typically composites of magnetic powder (i.e., iron, iron oxides, NdFeB, and others) and a soft, flexible matrix (i.e., polydimethylsiloxane and other silicone elastomers, polyurethane, and more). These artificial cilia are then actuated with an external magnetic field, produced by translating/rotating permanent magnets or electromagnets (Gauger et al. 2009; Shields et al. 2010; Gu et al. 2020; Hanasoge et al. 2020; Zhang et al. 2020, 2021; Islam et al. 2022; Peerlinck et al. 2023). Most studies involving these methods focus on low Reynolds number applications (e.g., microfluidics), producing fluid flow in viscous-dominated regimes (Islam et al. 2022; Sahadevan et al. 2022); thus, asymmetry is required to produce net flow. Furthermore, most magnetoactive artificial cilia are cylindrical in shape and achieve spatial asymmetry with 3D conical motions (Islam et al. 2022). Those that undergo 2D spatial asymmetry rely upon the cilia building elastic energy until they overcome the magnetic torque (Islam et al. 2022). However, the asymmetric nature of these 2D beating patterns can still be enhanced through other mechanisms, involving the optimization of propulsor shape and magnetic poling direction.

Here we present artificial cilia with plate-like shapes (Fig. 1c), inspired by the bundled cilia of ctenophores, in which we have passively encoded 2D spatial asymmetry. We use the same time-varying external magnetic field to actuate three different propulsor types—convex, concave, and flat (Fig. 1d-f)—across a range of beat frequencies and phase lag values. We show that our propulsors can produce beating patterns with similar asymmetry to real ctenophores (Fig. 1b; (Herrera-Amaya et al. 2021)) and represent a promising avenue for systematic investigation of the large parameter space governing the hydrodynamics of metachronal coordination. Here, using this new soft robotic platform, we explore the effects of asymmetry on fluid pumping performance across a range of Reynolds numbers.

## 2. Methods

### 2.1. Magnetic elastomer casting and magnetic poling

We constructed ctenophore-inspired magnetic-elastomer propulsors using methods similar to Gu et al. (Gu et al. 2020). Propulsor rows were digitally modeled then inverted to produce a 3D-printed mold (Fig. 2a). Propulsors were either curved or flat, with three row types: 1) convex, 2) concave, and 3) flat (where convexity/concavity is determined with respect to the power stroke direction, as shown in Fig. 3). Mold cavities for the individual propulsors were filled with a mixture of uncured silicone elastomer (Smooth-on Ecoflex$^{tm}$ 00-30) and neodymium alloy microparticles (NdFeB; median grain diameter ~40 μm) at a ratio of 1:1 by mass (14.4% by volume). The silicone-neodymium mixture was degassed in a vacuum chamber, which drew the silicone fully into the narrow voids of the mold. After curing, pure silicone elastomer was added on top of the propulsors (Fig. 2b), forming a nonmagnetic base for the propulsor row. Propulsor dimensions were 8 mm wide and tall and 0.9 mm thick; the row contained ten propulsors evenly spaced at 8.8 mm to avoid collisions. This length-to-height ratio falls within the range of real ctenophores, though relative thickness is considerably larger in these artificial ctenes compared to the biological model system (Afzelius 1961; Heimbichner Goebel et al. 2020). The artificial ctenes are approximately 8 times larger than the model system, but are dynamically scaled (via increasing the fluid kinematic viscosity) to match the typical Reynolds number of a beating ctene row (Tamm 2014; Heimbichner Goebel et al. 2020; Herrera-Amaya et al. 2021)

The flat propulsor row was magnetically poled before de-molding by placing the filled mold over the pole face of a Vibrating Sample Magnetometer (Lakeshore 8600), set to produce a uniform, unidirectional magnetic field at 1 Tesla. The curved propulsors, after curing and de-molding, were placed into a secondary mold that straightened them and tilted their bases 60 degrees (Fig. 2c). While in the secondary mold, these elastomers were placed over the electromagnet face under the same conditions. This configuration allowed the propulsors to be magnetically poled under strain, changing the magnetization direction along their lengths —the magnetic domains are vertical while in the strained (straight) shape (Fig. 2c), but when both the elastic forces (secondary mold) and electromagnetic forces (electromagnet) are removed, the propulsor returns to its curved shape (Fig. 2d). This procedure enables asymmetric motion under a time-varying external magnetic field, as described in the following sections. This two-part casting/poling technique is similar to that presented in a different context by Lin et al (Lin et al. 2023).

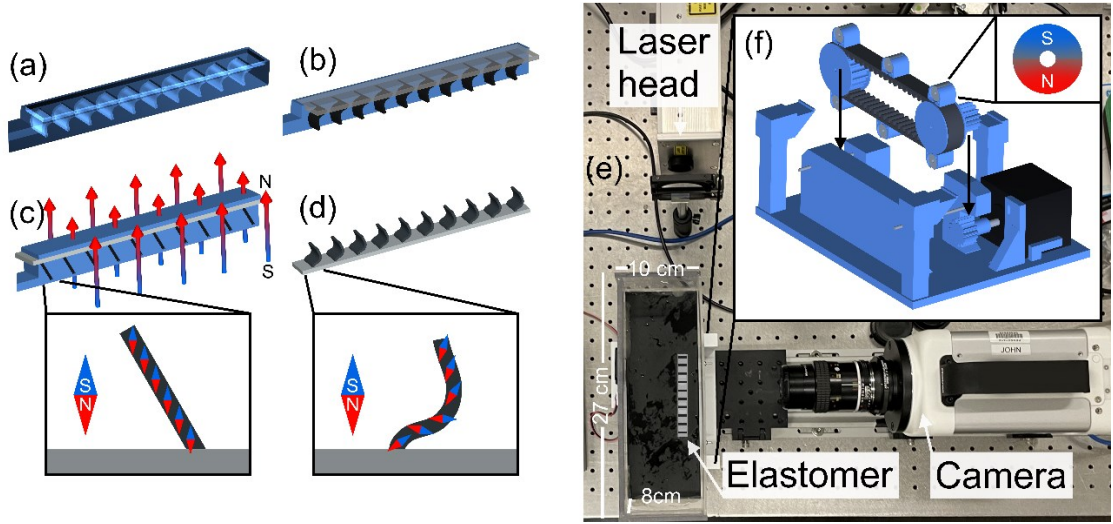

Figure 2: (a) 3D-printed mold used to cast ctenophore-inspired propulsors. (b) Cross-sectional view of the mold showing the propulsors (magnetic silicone elastomer; black), and the substrate (pure silicone elastomer; transparent gray). (c) Secondary mold used to straighten the curved propulsors during magnetic poling. Arrows denote the direction of the 1 Tesla magnetic field used to pole the samples. Bicolor diamonds indicate the directionality of magnetic domains along the length of the propulsor. (d) Demolded row of propulsors after magnetization; in the absence of an external magnetic field, magnetic domains reorient as shown in inset. (e) Experimental setup for Particle Image Velocimetry (PIV) showing relative locations of the infrared laser, high-speed camera, and acrylic tank (10 x 8 x 27cm width/depth/length) containing the ctenophore-inspired propulsor row (centerline 2.5 cm from tank wall). (f) Schematic of the 3D-printed apparatus, timing belt, and motor underneath the acrylic tank. This device translates diametrically poled cylinder magnets under the magnetic elastomer to produce a time-varying magnetic field.

## 2.2. Experimental setup

Each magnetized propulsor row was fixed to the bottom of a rectangular acrylic tank (Fig. 2e); the tank floor was recessed underneath the substrate, reducing the distance between the propulsors and the underlying actuating magnets to 1.5 mm (Fig. 2f). A 3D-printed platform (Fig. 2f) suspended the acrylic tank above translating diametrically poled cylinder magnets (diameter 9.525 mm, K&J Magnetics, Inc., model R6036DIA, grade N42) attached to a 30 cm timing belt (following (Zhang et al. 2020)). The moving magnets impose a time-varying magnetic field, such that the propulsors beat in an antiplectic wave (i.e., wave propagates in the opposite direction of the power stroke; see supplemental footage; doi.org/10.5281/zenodo.11640274). The south pole of each magnet points outwards, straightening the propulsors in the near-field and allowing their elasticity to rebound to the original shape in the far-field (see Fig. 3).

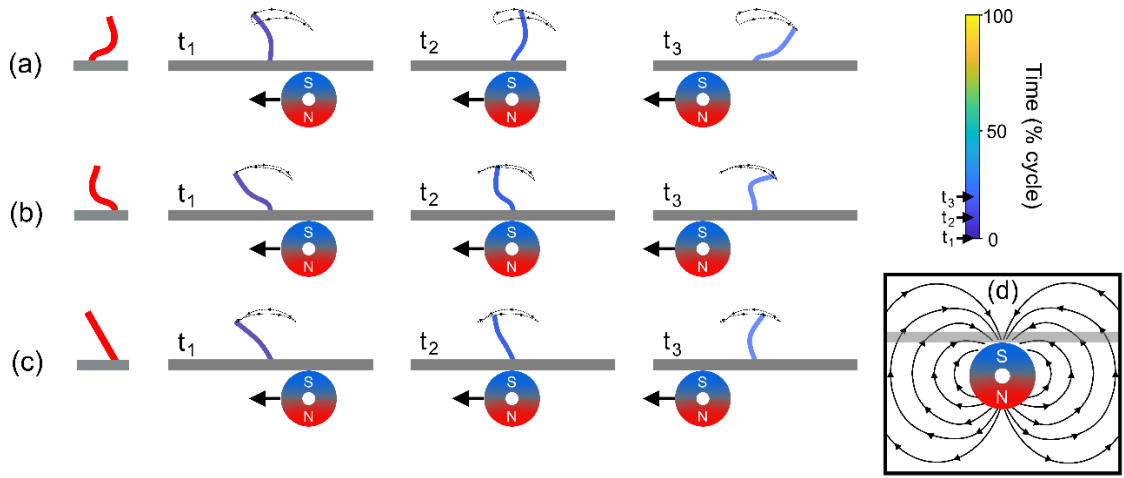

Figure 3: Schematic of magnet positions underlying the three propulsor types, (a) convex, (b) concave, and (c) flat, at three timesteps ($t_1$ = magnet approaching, $t_2$ = magnet underneath propulsor, and $t_3$ = magnet leaving). Red propulsors show the undeformed shape in the absence of a magnetic field. (d) Depiction of magnetic flux lines surrounding the diametrically polarized magnet. Note that the (magnetized) propulsors themselves also influence the overall magnetic field.

    The spacing of the magnets on the timing belt controls the wavelength $\lambda$ of the antiplectic wave (Fig. 4). The speed of the timing belt is equal to the wave speed $c$. The frequency is therefore $f = c/\lambda$. We are interested in varying both phase lag and frequency, where phase lag is expressed as percentage of the beat cycle (such that a phase lag of 25% indicates that each ctene lags its neighbor by a time interval that is 25% of the overall wave period). The spacing between propulsors $\delta$ is fixed; the phase lag $PL = 1/n_{row}$ where $n_{row}$ is the number of propulsors per wavelength $\lambda$. We see that $n_{row} = \lambda/\delta$ and thus, $PL = \delta/\lambda$ – this shows that while beat frequency is controlled by both wavelength (magnet spacing) and wave speed (belt speed), phase lag is controlled exclusively by wavelength (magnet spacing). Doubling wavelength halves phase lag; thus, preserving the frequency for two different phase lags requires changing the belt speed (for example, if $\lambda_1 = 2\lambda_2$ then $c_1 = c_2/2$).

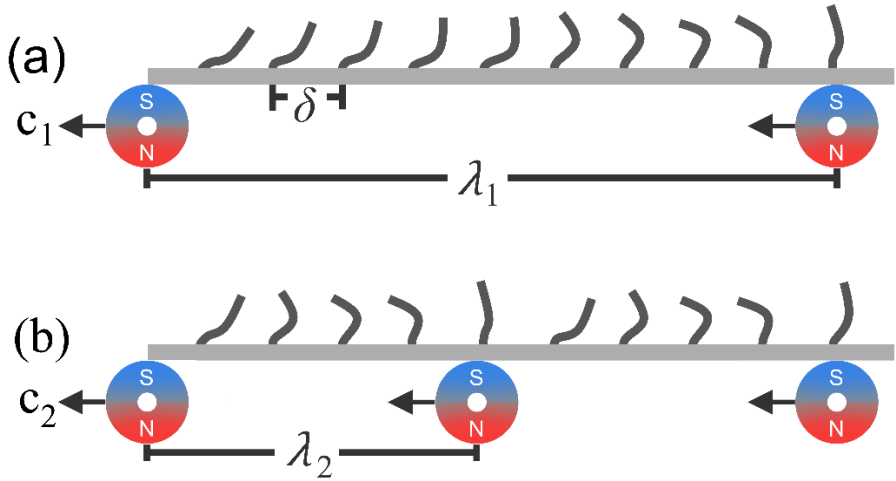

Figure 4: Relationship between magnet spacing (i.e., wavelength, $\lambda$), wave speed c, phase lag PL, and propulsor spacing $\delta$. (a) PL = 8.8%, $\lambda_1$ = 10 cm. (b) PL = 17.6%, $\lambda_2$ = 5 cm. To match frequency, the timing belt must be driven at $c_2 = 2c_1$.

We chose two target phase lag values within a biologically relevant range for real ctenophores (Barlow and Sleigh 1993), 8.8% and 17.6%. These values correspond to one and two magnets per total propulsor row length, respectively. For each phase lag, we actuated the propulsors at 4 and 8 Hz. The experiments at 8 Hz beat frequency and 8.8% phase lag used a 12V DC motor to achieve the high belt speeds. For all other experiments, a NEMA 17 stepper motor was used to drive the timing belt. To ensure dynamic similarity with the biological model system, experiments were conducted in a 70% glycerol-water mixture (kinematic viscosity $\nu = 2.37 \cdot 10^{-5}$ m$^2$/s, measured with an AMETEK Brookfield viscometer at 21°C, model LVDVE115), yielding intermediate Reynolds numbers consistent with active ctenes (~1<Re<150).

### 2.3. Particle Image Velocimetry (PIV)

We used planar Particle Image Velocimetry (2D2C PIV) to characterize fluid flow above the propulsors at varying frequency and phase lag. An infrared laser (Oxford Lasers FireFLY, wavelength 808 ± 3 nm) illuminated 11 μm glass microspheres (Sphericel, Potters Industries) in a plane intersecting the midline of the propulsor row. The laser aperture was 24 cm from the midpoint of the propulsor row; the beam was expanded both with internal optics and an additional plano-concave cylinder lens (Thorlabs; focal length = -19 mm; radius = -9.8; length = 21 mm; height = 19 mm), placed 2.54 cm from the aperture. A high-speed camera (Photron FASTCAM NOVA R5), fitted with a 55mm macro lens (Micro-Nikkor, Nikon), viewed the light sheet orthogonally with its sensor 32 cm from the light sheet, leading to a pixel size of 18.19 μm/px in the produced images. All footage was recorded at 1000 frames per second at 3840 by 2160 px resolution, with laser pulses of 100 μs externally synchronized via the camera using the leading edge of each frame. Five cycles were recorded for five propulsors from the middle of the row for each combination of propulsor type, phase lag, and beat frequency. While eight of the ten propulsors were visible in the camera's field of view, the analysis region was limited to the field above the central five propulsors to avoid potential edge artifacts (Fig. S1). The propulsors actuated for at least one minute prior to filming to avoid transient start-up conditions. All footage was processed using DaVis 10.2.1 (LaVision GmbH, Goettingen, Germany), with a final subwindow size of 64 x 64 pixels with 50% overlap leading to a vector spacing of 0.582 mm, with the area around the propulsors masked to avoid boundary artifacts (see supplemental footage; doi.org/10.5281/zenodo.11640274).

## 2.4. Propulsor kinematics

The tips and bases of five adjacent propulsors were tracked for five cycles, using DLTdv8 (Hedrick 2008). The framerate was down-sampled to 500 fps for the 4 Hz experiments and kept at 1000 fps for the 8 Hz experiments to equalize the number of tracked frames per cycle. Stroke angle $\beta(t)$ and stroke amplitude $\Phi$ (range of stroke angle) were computed from the vector between tip and base. Maximum tip velocity $U_{max}$ was used as the characteristic velocity for the Reynolds number, such that

$$Re = \frac{U_{max} L}{\nu} \quad (1)$$

where L is the propulsor length (8 mm) and $\nu$ is the fluid kinematic viscosity.

Phase lag was computed directly as the time lag between the start of a power stroke for neighboring propulsors, normalized by the cycle duration. It can also be computed a priori from geometric variables as $\delta/\lambda$, as discussed previously.

Spatial asymmetry ($Sa$) characterizes the difference in swept area between the power stroke and recovery stroke (Herrera-Amaya et al. 2021). It is defined as

$$Sa = \frac{A_e}{A_0} \quad (2)$$

where $A_e$ is the area defined by the integration of the propulsor tip's path over one beat cycle (Fig. 1b), and $A_0$ is the largest possible area of an ellipse inscribed within the reachable space of the propulsor (i.e., a half-circle with a radius equal to the propulsor length; Fig. 1b; $A_0 = 0.77\pi L/2$; (Herrera-Amaya et al. 2021)). $Sa$ approaches one for beat cycles with high spatial asymmetry, and zero for propulsors moving symmetrically (i.e., reciprocal beating). $Sa$ may be negative, implying that the flow-normal area is lower during the power stroke vs. the recovery stroke.

Temporal asymmetry ($Ta$) characterizes the difference between the duration of the power vs. recovery stroke (Gauger et al. 2009; Herrera-Amaya et al. 2021), such that

$$Ta = \frac{t_r - t_p}{t_r + t_p} \quad (3)$$

where $t_r$ is the recovery stroke duration and $t_p$ is the power stroke duration. $Ta$ approaches zero when the power stroke and recovery stroke have equal durations (symmetric in time), and approaches one when the power stroke is infinitely fast (asymmetric in time). $Ta$ may also be negative.

Means and standard deviations for each of these kinematic properties are listed in Table S1 and the raw values (for each cycle and each propulsor) are located in Table S2.

## 2.5. Characterizing pumping performance

Cycle-averaged momentum was computed from the velocity field as the spatiotemporal average of instantaneous measurements throughout the region of interest (Fig. S1) over one beat cycle. The positive horizontal direction coincides with the power stroke direction, while the positive vertical direction is upwards and aligned with gravity (Fig. S1). The horizontal and vertical cycle-averaged momentum are defined as:

$$\overline{\rho u} = \frac{\rho}{n_u n_v n_f} \sum_{i=1}^{n_u} \sum_{j=1}^{n_v} \sum_{k=1}^{n_f} u(i,j,k) \quad (4)$$

$$\overline{\rho v} = \frac{\rho}{n_u n_v n_f} \sum_{i=1}^{n_u} \sum_{j=1}^{n_v} \sum_{k=1}^{n_f} v(i,j,k) \quad (5)$$

where $\rho$ is the fluid density (1189.1 kg/m³), $n_u$ and $n_v$ are the number of PIV interrogation windows in the horizontal and vertical directions, $n_f$ is the number of frames in each cycle, and $u$ and $v$ are the horizontal and vertical components of velocity. The magnitude of cycle-averaged momentum ($\overline{\rho U}$) and its angle relative to the substrate ($\bar{\theta}$) were computed as:

$$\overline{\rho U} = \sqrt{\overline{\rho u}^2 + \overline{\rho v}^2} \quad (6)$$

$$\bar{\theta} = \tan^{-1}\left(\frac{\overline{\rho v}}{\overline{\rho u}}\right) \quad (7)$$

All means, standard deviations, and per-cycle values related to cycle-averaged momentum are listed in Table S3.

## 3. Results

### 3.1. Propulsor kinematics

Our soft robotic platform allows propulsor beat frequency $f$ and phase lag $PL$ to be prescribed by varying the speed of the timing belt and the spacing of the actuating magnets (Figs. 2f, 4). For each combination of $f$ and $PL$, our three propulsor types experience the same actuating magnetic field but yield different kinematics. The three types are referred to as 1) convex, 2) concave, and 3) flat (where the convex/concave types are identical but rotated 180 degrees about the vertical axis). Each type is actuated at a low and high frequency ($f = 4$ and 8 Hz), and a low and high phase lag ($PL = 8.8\%$ and 17.6%). When the actuating magnets are far from a given propulsor, it assumes its undeformed shape as determined by initial molding (Fig. 2). When a magnet translates directly under a propulsor, magnetic flux lines enter the tank orthogonal to the surface. Far from the pole axis of the actuating magnets, these flux lines point in the opposite direction, and they are oblique in regions in between (Fig. 3d). Consequently, the magnetic field changes in strength and direction over time, pushing then pulling the propulsors as the actuating magnets pass underneath (Fig. 3). Each propulsor experiences both magnetic forcing (to align with the magnetic field according to the magnetization direction encoded during poling of the elastomers; Fig. 2c,d) and elastic forcing (to assume the initially-molded shape; Fig. 2a,b). When a magnet passes directly underneath a propulsor, the dominant forcing is magnetic; the propulsor bends until elastic forcing can overcome the magnetic forcing, which weakens as the magnets translate away. Eventually, the propulsor returns to the shape in which it was cast (until the next magnet approaches). Fluid drag also acts on the propulsors during both the power and recovery stroke.

Stroke amplitude $\Phi$ (Fig. 1d) ranges from ~70 to ~100 degrees across the tested frequencies and phase lags (Table S1). In general, $\Phi$ decreases at higher frequencies because the timescale of the magnetic forcing decreases (so that elastic forcing is more dominant). The combination of magnetic and elastic forcing also produces temporal asymmetry. For a given experiment, the convex propulsors typically assume lower stroke angles just after the end of the power stroke. That is, they spend more time bent forward before being reset by the recovery stroke. In contrast, the concave propulsors more quickly return to their original orientation after the power stroke, and the flat propulsors undergo near-sinusoidal oscillations in stroke angle (Figs. 5 and S2-4; Tables S1 and S4).

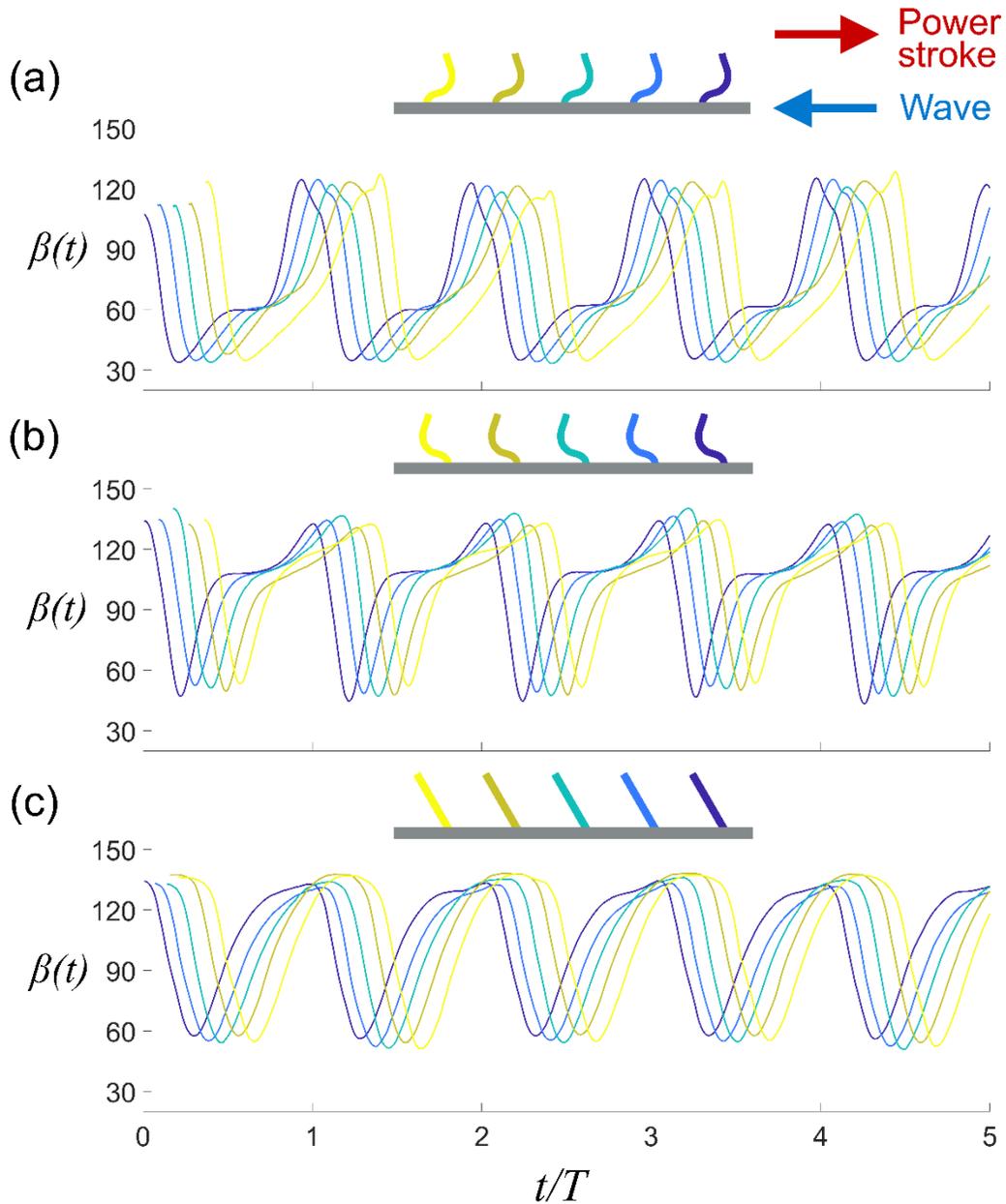

Figure 5: Stroke angle $\beta(t)$ of five neighboring propulsors over five beat cycles ($f = 4$ Hz, $PL = 8.8\%$); kinematics differ between (a) convex, (b) concave, and (c) flat propulsor types. Stroke angles for all other experiments can be found in Figs. S2-4. Arrows denote the opposing directions of the power stroke and metachronal wave (antiplectic metachrony).

Both curved propulsor types consistently have higher spatial asymmetry (Fig. 6a) compared to the flat case. Compared to their cast shape, the curved propulsors (Fig. 1d,e) straighten out more just before the actuating magnet passes underneath (Fig. 2c). The convex propulsors generally assume a straighter shape during the power stroke with a more bent recovery stroke Fig. 1d), which increases $Sa$ for most experiments (Table S1). In the convex configuration, the magnetic torque unfolds the curved propulsors, aligning the domains to oblique magnetic flux lines as the magnet passes underneath (Figs. 2d, 3a,d). However, in the concave configuration (Fig. 1e), propulsors retain comparatively more curvature during the power stroke and thus have a lower $Sa$ (except for $f = 4$ Hz and $PL = 17.6\%$, see Fig. 4). The poling orientation of the propulsors relative to the magnetic field causes these differences in bending. When the magnet passes underneath, the convex propulsors align to the magnetic field in a way that works against the curvature of the cast shape. That is, the north poles of the domains follow a curved surface that has opposite curvature to the elliptical magnetic flux lines (Figs. 2d and 3d). Conversely, the concave propulsors experience comparatively more folding during the power stroke because the orientation of the magnetic domains is mirrored about the vertical axis. The flat propulsors undergo nearly reciprocal motion but bend more during the power stroke due to fluid drag, producing negative spatial asymmetry (Fig. 1f, Fig. 4). The curved propulsors also have higher temporal asymmetry due to their relatively shorter power stroke durations (Fig. 6b), likely from the complex interplay between magnetic, elastic and fluid forcing. Though, further investigation and modeling is required to better understand the relative contributions of each mechanism influencing overall propulsor kinematics.

Means and standard deviations for all kinematic properties were computed for five propulsors over five cycles (Table S1; for raw values see Table S2).

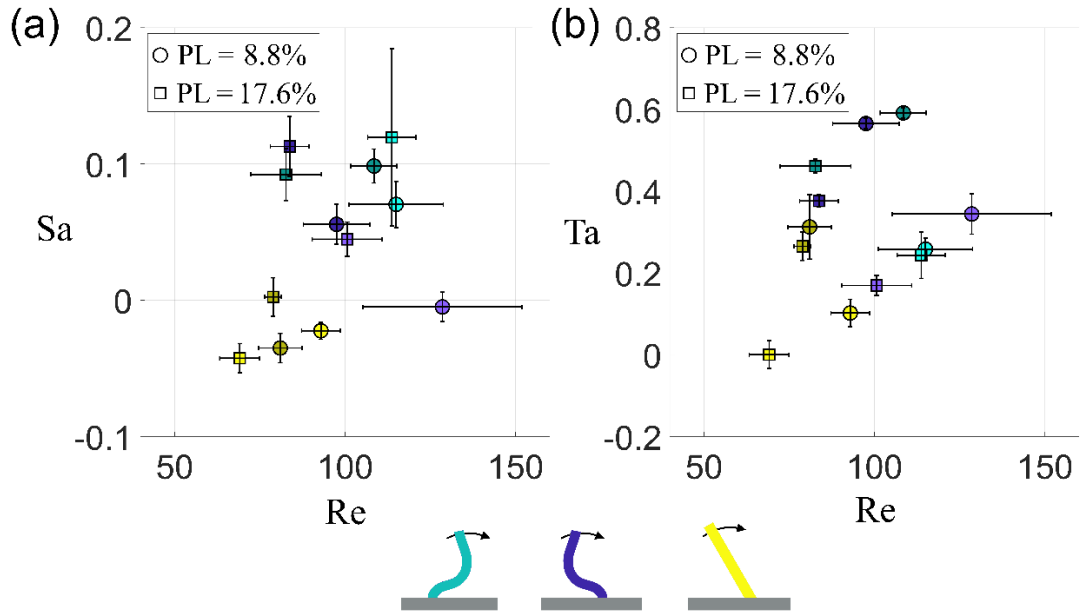

Figure 6: a) Spatial asymmetry ($Sa$) vs Reynolds number ($Re$; based on $U_{max}$). b) Temporal asymmetry ($Ta$) vs $Re$. Color represents propulsor type (teal = convex; purple = concave; yellow = flat); lighter colors represent higher frequency ($f = 8$Hz) and darker colors represent lower frequency ($f = 4$Hz). Marker shape represents phase lag. Arrows on propulsors indicate the power stroke direction.

**3.2. Fluid pumping performance**

We used cycle-averaged momentum (Fig. 7) to measure pumping performance. It is calculated as a spatiotemporal average of $\rho u(\vec{x}, t)$ and $\rho v(\vec{x}, t)$ across the entire region of interest above five propulsor tips (Fig. S1), and throughout the duration of a single beat cycle (Equations 4-6). The resulting values of $\overline{\rho u}$ and $\overline{\rho v}$ were averaged over five beat cycles (raw values, means and standard deviations reported in Tables S3). The horizontal cycle-averaged momentum ($\overline{\rho u}$) is most relevant for pumping performance, characterizing the net fluid motion parallel to the substrate. Both curved propulsor shapes outperform the flat ones (Fig. 7b), though the difference is most extreme for the convex case. In general, higher propulsor speeds impart more momentum into the overlying fluid; thus, tip speed (and therefore Reynolds number) should be considered when comparing pumping performance between experiments. We note that tip speed depends on the unique combination of magnetic, elastic, and fluid forcing for a given propulsor type, and thus actuating two different propulsor types at the same frequency may produce different Reynolds numbers. At a given frequency and phase lag, $Re$ is generally lower for the flat propulsors relative to the two curved types, which have comparable $Re$. However, despite similarities in $Re$, the convex propulsors generate 2.2 to 3.8 times the horizontal cycle-averaged momentum compared to the concave case (Fig. 7b; Table S3). Additionally, the concave propulsors generally have a larger $\overline{\rho u}$ relative to $\overline{\rho v}$, indicating that input energy is directed more towards producing substrate-parallel (vs. substrate-orthogonal) flow (Fig. 8 and Table S3). The flat propulsors consistently produce net flows at (average) higher angles from the substrate, and in one case (8 Hz, 17.6% phase lag) produces net upwards and backwards flow (Fig. 8 and Table S3).

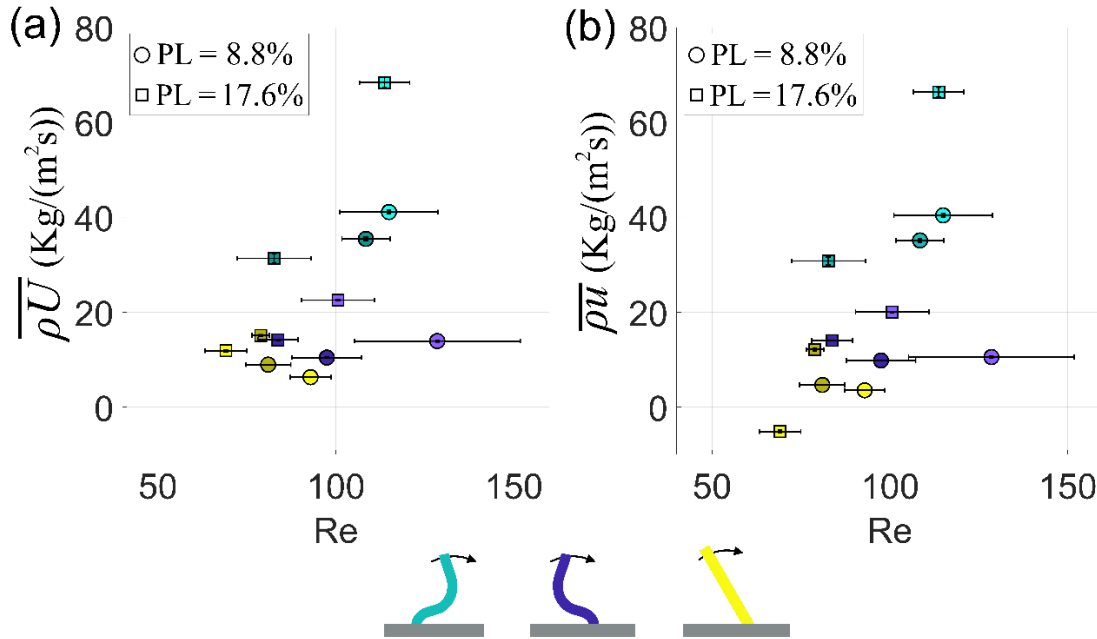

Figure 7: Cycle-averaged momentum vs $Re$. (a) Overall magnitude of cycle-averaged momentum ($\overline{\rho U}$). (b) Horizontal cycle-averaged momentum ($\overline{\rho u}$). Color represents propulsor type (teal = convex; purple = concave; yellow = flat); lighter colors represent higher frequency

($f = 8$Hz) and darker colors represent lower frequency ($f = 4$Hz). Marker shape represents phase lag. Arrows on propulsors indicate the power stroke direction.

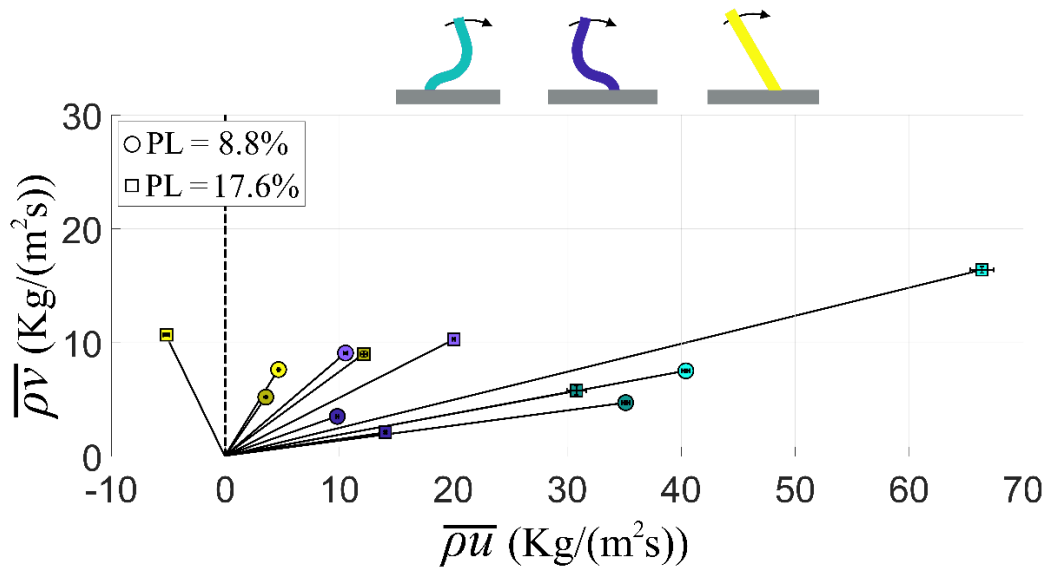

Figure 8: Vertical cycle-averaged momentum ($\overline{\rho v}$) vs horizontal ($\overline{\rho u}$). Lines from the origin to each point represent the angle from the horizontal of the average flow (for momentum angle $\bar{\theta}$, see Table S3), with dashed line showing the vertical. Color represents propulsor type (teal = convex; purple = concave; yellow = flat); lighter colors represent higher frequency ($f = 8$Hz) and darker colors represent lower frequency ($f = 4$Hz). Arrows on propulsors indicate the power stroke direction.

    The phase-averaged velocity (calculated by averaging the flow fields at the same time in the cycle, for five beat cycles; for videos, see supplementary footage (doi.org/10.5281/zenodo.11640274)) yields insight into the spatial variation of the velocity and vorticity field. Figure 9 is a snapshot of the phase-averaged horizontal velocity at the moment the center propulsor begins its power stroke-. Through the entire beat cycle, the convex propulsors produce steady horizontal flow above the propulsor row, with little impeded (low $u$) or backwardly directed flow (negative $u$). The concave propulsors generate similar horizontal flow above the propulsor row; however, during each propulsor's power stroke, transient regions of backwards flow appear due to fluid filling the low pressure regions behind the propulsors. This behavior weakens the horizontal flow above the propulsor (and therefore the overall pumping performance). The flat propulsors produce lower velocity flows and generate the most backward flow compared to the other propulsor shapes. Backward-flow regions briefly occur above the propulsors during the power stroke, but the recovery stroke is responsible for generating substantially larger regions of backwards flow, producing a coherent negative shear layer, such that the velocity field contains alternating layers of positive and negative $u$ (see supplementary footage (doi.org/10.5281/zenodo.11640274)).

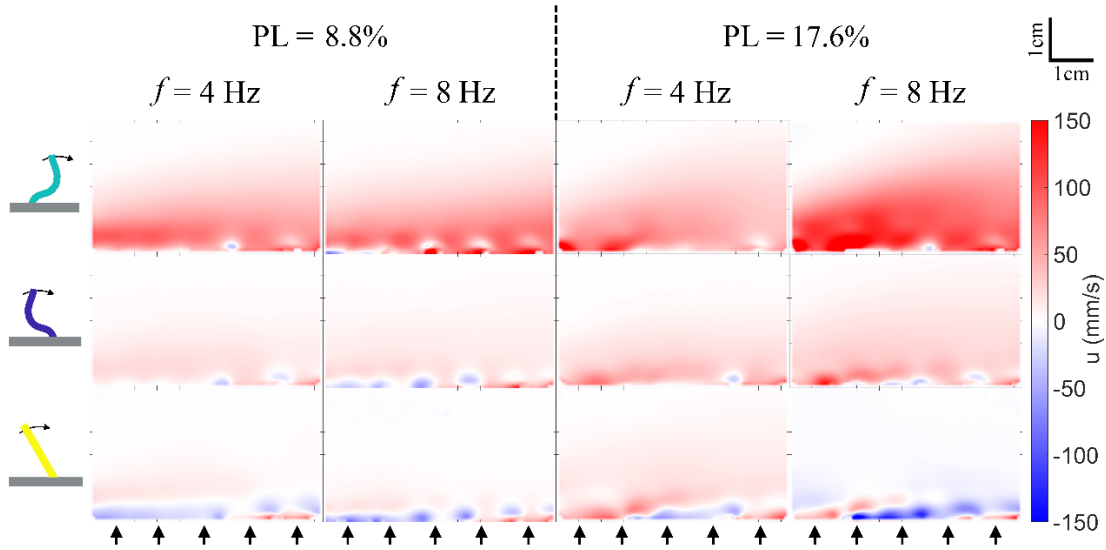

Figure 9: Phase-averaged horizontal velocity ($u$), shown at the instant the center propulsor begins its power stroke. Each panel shows a distinct combination of frequency $f$, phase lag PL and propulsor type (top row = convex; middle = concave; bottom = flat). Black arrows at the bottom of each panel denote the approximate location of each propulsor base (below the region of interest). For videos of the entire beat cycle see supplementary footage (doi.org/10.5281/zenodo.11640274).

From this velocity field, we calculate the out-of-plane vorticity $\omega_z \equiv \frac{\partial v}{\partial x} - \frac{\partial u}{\partial y}$. We calculate a phase-averaged vorticity field to examine consistent flow structures that develop during the power-recovery cycle of each paddle (Fig. 10; supplementary footage, doi.org/10.5281/zenodo.11640274). We observe similarity between the artificial propulsors and the biological model system, including the development of discrete positive vortices (as shown in red) during the power stroke compared to a continuous band of negative vorticity (as shown in blue) during the recovery stroke. The discrete positive vortices are associated with thrust generation; the negative band of vorticity generates drag (negative horizontal flow). The flexibility of the propulsors, combined with the spatial asymmetry exhibited by the two curved cases, allows the propulsor tips to be relatively further apart during the power stroke vs. the recovery stroke. This spatial arrangement enhances the positive vortices (created during the power stroke) and weakens the negative vortices (created during the recovery stroke), which increases the net thrust generated over a single cycle. This arrangement and fluid dynamic mechanism is similar to what occurs in the biological model system (Lionetti et al. 2023). However, a more detailed exploration of vorticity generation and its dependence on propulsor kinematics and spacing is needed to draw further conclusions.

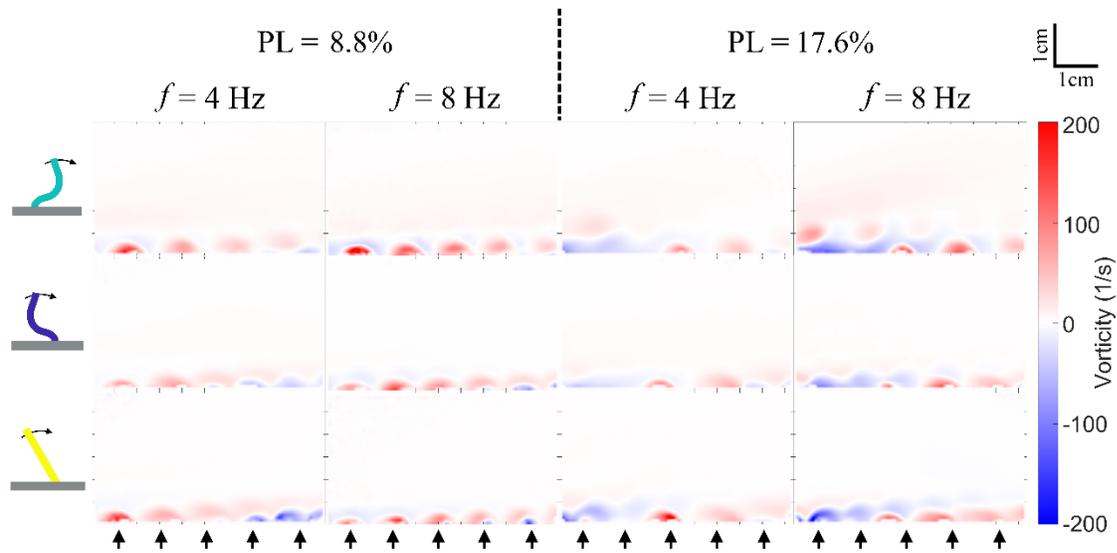

Figure 10: Phase-averaged vorticity $\omega_z$, shown at the instant the center propulsor begins its power stroke. Each panel shows a distinct combination of frequency $f$, phase lag $\Phi$, and propulsor type (top row = convex; middle = concave; bottom = flat). Black arrows at the bottom of each panel denote the approximate location of each propulsor base (below the region of interest). For videos of the entire beat cycle see supplementary footage (doi.org/10.5281/zenodo.11640274).

## 4. Discussion and conclusion

Elastic, magnetic, and fluid forcing all play important roles in the kinematics of propulsors made from magnetoactive silicone elastomers. Using a secondary mold to magnetically pole the curved propulsors under elastic strain (Fig. 2c) produces a mismatch between the magnetization direction along the length of the propulsor and the cast shape (Fig. 2c,d): when the magnetic forcing is strong (during the power stroke), the propulsor will straighten, aligning with the orientation in which it was poled. During the recovery stroke, when the actuating magnetic field is weaker, elastic forcing returns the propulsors to their initially curved shape. This technique can be used to passively encode 2D spatiotemporal asymmetry in paddle-shaped propulsors. Due to the combination of elastic and magnetic forcing, the curved propulsors extend during the power stroke and retract during the recovery stroke, increasing spatial asymmetry ($Sa$). Additionally, the duration of the power stroke relative to the recovery stroke is shorter, improving temporal asymmetry ($Ta$). This occurs because the timescale of elastic forcing is generally longer than the timescale of magnetic forcing—that is, the actuating magnetic field is present for a time $\tau_m$, and after it passes the propulsor requires time $\tau_e$ to return to its undeformed shape (where $\tau_e > \tau_m$).

These soft robotic propulsor rows offer insights into 1) how metachronally coordinated appendages interact with fluids, 2) the functional morphology of meso-scale aquatic organisms, and 3) engineering solutions to the design of aquatic soft robots and pumping devices.

## 4.1. Relationships between propulsor kinematics and pumping performance

Both propulsor kinematics and propulsor speed (Table S1) control the strength of the flows produced during metachronal beating. Even though beat frequency $f$ and phase lag $PL$ are prescribed, the tip speed of the propulsors can differ due to the coupling of experimental parameters and differences in kinematics. As discussed in section 2.2, the number of magnets on the timing belt must be altered to adjust phase lag. Doubling $PL$ while holding $f$ constant requires the timing belt to move at half the speed, which increases the magnetic forcing timescale $\tau_m$. This not only lowers the tip velocity of each propulsor, but also decreases $\tau_m$ with respect to the elastic timescale $\tau_e$, thereby decreasing temporal asymmetry ($Ta$). Additionally, kinematics may differ slightly for any given propulsor shape depending on the phase lag and beat frequency, since the relationship between $\tau_m$ and $\tau_e$ will affect fluid-structure interactions. Consequently, for a given $PL$, increasing $f$ usually reduces spatial asymmetry and always reduces temporal asymmetry (Fig. 6; Table S1). These relationships are complex, underscoring the potential for fine-tuning of magnetic and elastic forcing to produce a set of desired propulsor kinematics.

Despite the coupling between parameters, our platform reveals interesting relationships between propulsor kinematics and produced flows. Of the three propulsor shapes (convex, concave, and flat), the convex propulsors produce considerably higher fluid flow (net cycle-averaged momentum; Fig 7a). Flows produced by the convex propulsors were also directed more horizontally (parallel to the substrate; Fig. 7b, Fig. 8), which is relevant for swimming and pumping (though lift generation is also important for hovering in organisms that use metachronal coordination (Murphy et al. 2011; Ford and Santhanakrishnan 2021)). The flat propulsors produce the weakest flows due to their low and/or negative spatial asymmetry and lower temporal asymmetry (for all combinations of beat frequency and phase lag; Fig. 6). The reduction in pumping performance can also be attributed to lower propulsor speeds (as shown by lower $Re$) for the flat propulsors (Fig. 6, Fig. 7). However, the convex propulsors produce considerably more cycle-averaged momentum compared to the concave shape, despite having similar $Re$ (Fig. 7, Table S3). Additionally, the convex propulsors often have slightly lower stroke amplitude Φ (Table S1) compared to the other shapes, yet do not experience any reduction in flow. This is in part due to the kinematics of each propulsor at the end of the power stroke. The convex propulsors slow down just after completion of the power stroke and prior to commencing the recovery stroke (wider troughs on Figs. 5 and S4). This "resting" phase, which is similar to what we observe in real ctenophores (Herrera-Amaya et al. 2021), allows moving fluid to "coast" above the propulsors prior to the disruptive onset of the recovery stroke. This tendency is only observable (or useful) in intermediate-to-high $Re$ systems, and cannot occur in low-$Re$ systems in which viscous forces dominate over inertial forces. It is also not easily incorporated into the concept of temporal asymmetry, which requires more nuanced exploration in both this system and similar biological models. These experiments demonstrate that nuanced differences in propulsor kinematics can significantly influence the generated flows, particularly around the viscous-to-inertial transition regime (intermediate $Re$).

An examination of the phase-averaged velocity and vorticity fields (Figs. 9 and 10) reveal that shifts in propulsor kinematics can significantly change the overall performance characteristics of the system. The convex propulsors produce the most consistent positive flow velocity, indicating that they are the most efficient producers of thrust; by contrast, the flat propulsors move fluid both positively and negatively, leading to reduced efficiency. The convex propulsors produce discrete regions of positive (thrust-producing) vorticity and weaker, smaller bands of negative (drag-producing vorticity), as seen in real ctenophores, whereas the flat propulsors produce equivalent positive vs negative vorticity. Further exploration of how differences in propulsor shape throughout the beating cycle contribute to the overall performance of the array is needed.

**4.2. Bioinspired metachronal coordination**

Many marine animals generate flows within a transitional physical regime where both viscous and inertial effects are important (intermediate $Re$). Within this $Re$ range, natural selection has produced numerous variations on metachronally coordinated paddle-like propulsors (e.g., the ctenes of ctenophores, pleopods of crustaceans, parapodia of polychaetes, etc.) rather than appendages with the cylindrical shapes of motile cilia and flagella, which are common at low $Re$. These paddle-shaped propulsors beat with 2D spatial asymmetry rather than rotating conical motions or whip-like motions (Craig and Okubo 1990; Lim and DeMont 2009; Murphy et al. 2011; Tamm 2014; Ford et al. 2019; Daniels et al. 2021; Garayev and Murphy 2021; Zhang et al. 2021; Peerlinck et al. 2023). At intermediate $Re$, the drag-based metachronal paddling of these propulsor shapes can offer increased performance in acceleration, braking, and turning compared to other types of locomotion, like lift-based swimming (Vogel 1994, 2013; Walker and Westneat 2000; Byron et al. 2021). The morphological convergence of metachronally coordinated appendages—and related kinematic properties—emphasize their potential value for the design of bioinspired swimming robots and/or pumping devices at the millimeter scale. The technique we present here to encode these motions in artificial propulsors offers several benefits. First, it presents a mechanically simple way to coordinate many appendages without independent motors or geartrains. Second, the actuating field has no physical connection to the propulsors themselves—a boon for free-swimming robots that must be watertight to maintain proper hydrostatics and to protect on-board electronics and sensors. Third, the propulsor rows are modular and can be easily swapped to adjust for desired kinematics; the technique of magnetically poling the propulsors under elastic strain can be adapted to encode other types of spatial asymmetry. For example, many paddle-shaped propulsors splay laterally during the power stroke to increase their area relative to the recovery stroke (Wootton 1999; Kim and Gharib 2011; Heimbichner Goebel et al. 2020; Santos et al. 2023). These kinematics could be achieved with a magnetically actuated, folding propulsor (e.g. with magnetoactive origami techniques; (Cowan and von Lockette 2017; Lin et al. 2023)). Additionally, the magnetic-elastic approaches described here can be extended with composite materials to achieve the desired asymmetric bending of the propulsors.

The results presented here demonstrate a technological solution to achieve 2D spatiotemporal asymmetry in magnetoactive materials, while highlighting the importance of nuanced propulsor kinematics for swimming at the meso-scale. This insight is generalizable not only to swimming, but to feeding, respiration, and many other functional behaviors that require the efficient and effective movement of fluid. The large parameter space of metachronal rowing (and of coordinated flexible propulsors more generally) represents a rich and complex system that carries high potential for adaptation into new bioinspired devices and vehicles; our results here represent one path forward into this domain while emphasizing the importance of propulsor kinematics to the overall generated flows.

## Acknowledgements

We greatly appreciate the help of Dashiell Papula, Paris von Lockette, Subrata Ghosh, Sumanta Karan, and Bed Poudel for assistance with magnetically poling samples, and for discussions that considerably helped with magnetic elastomer prototyping and design.